\documentclass[12pt, twoside, a4paper]{article}
\usepackage{imm17_4}

\pdfoutput=1

\usepackage[pdftex, unicode]{hyperref}
\usepackage[pdftex]{graphicx}

\hypersetup{pdftitle={Analysis of the Time Structure of Synchronization in Multidimensional Chaotic Systems}}
\hypersetup{pdfauthor={A.\,V.\,Makarenko, Constructive Cybernetics Research Group}}
\hypersetup{pdfsubject={Nonlinear Dynamics}}
\hypersetup{pdfkeywords={chaos, T-synchronization, intermittency, time structure, synchronization patterns, desynchronization areas.}}

\newcommand{\MaincltStr}{arXiv.org (J.~Exp.~Theor.~Phys., 2015, Vol.~120, No.~5, pp.~912--921)}

\newcommand{\Leftclt}{A.\,V.\,Makarenko}

\newcommand{\Rightclt}{Analysis of the Time Structure of Synchronization\ldots}

\newcommand{\UDK}{MSC: 37M10, 37M20, 34C28, 34C15, 34D06;
\\ PACS: 02.70.-c, 05.45.-a, 05.45.Tp, 05.45.Xt.}

\begin{document}

\Mainclt 

\begin{center}
\Large{\bf Analysis of the Time Structure of Synchronization\\
in Multidimensional Chaotic Systems}\\[2ex]
\end{center}

\begin{center}
\large\bf{A.\,V.\,Makarenko}\supit{a,}\supit{b,}
\footnote{E-mail: avm.science@mail.ru}\\[2ex]
\end{center}

\begin{center}
\supit{a}\normalsize{Constructive Cybernetics Research Group}
\\
\normalsize{P.O.Box~560, Moscow, 101000 Russia}\\[3ex]

\supit{b}
\normalsize{Institute of Control Sciences, Russian Academy of Sciences}
\\
\normalsize{ul.~Profsoyuznaya~65, Moscow, 117977 Russia}\\[3ex]
\end{center}

\small{Received October 28, 2014}
\begin{quote}\small
{\bf Abstract}. A new approach is proposed to the integrated analysis of the time structure of synchronization of multidimensional chaotic systems. The method allows one to diagnose and quantitatively evaluate the intermittency characteristics during synchronization of chaotic oscillations in the T-synchronization mode. A system of two identical logistic mappings with unidirectional coupling that operate in the developed chaos regime is analyzed. It is shown that the widely used approach, in which only synchronization patterns are subjected to analysis while desynchronization areas are considered as a background signal and removed from analysis, should be considered as methodologically incomplete.
\end{quote}


\begin{Keyworden}
chaos, T-synchronization, intermittency, time structure, synchronization patterns, desynchronization areas.
\end{Keyworden}


\setcounter{equation}{0}
\setcounter{lem}{0}
\setcounter{teo}{0}



\section{Introduction}

Synchronization is one of the fundamental concepts of the theory of nonlinear dynamics and chaos
theory. This phenomenon is widespread in nature, science, engineering, and society [1]. One of important
manifestations of this phenomenon is the synchronization of chaotic oscillations, which was experimentally observed in various physical applications (see~\cite{bib:book_Pikovsky_2001, bib:article_Boccaletti_PhysRep_2002_366, bib:article_Argonov_JETPL_2004_80, bib:article_Kuznetsov_UFN_2011_2, bib:article_Napartovich_JETP_1999_5} and references therein) such as radio oscillators, mechanical systems, lasers, electrochemical oscillators, plasma and gas discharge, and quantum systems. The study of this phenomenon is also very important from the viewpoint of its application to information transmission~\cite{bib:article_Cuomo_PhysRevLett_1993_71}, cryptographic coding~\cite{bib:article_Larger_Physique_2004_5} with the use of deterministic chaotic oscillations, and quantum computation~\cite{bib:article_Argonov_JETPL_2004_80, bib:article_Planat_Neuroquantology_2004_2}.

There are several types of synchronization of chaotic oscillations~\cite{bib:article_Boccaletti_PhysRep_2002_366}:
generalized synchronization~\cite{bib:article_Abarbanel_PhysRevE_1996_53},
complete synchronization~\cite{bib:article_Pecora_PhysRevLett_1990_64},
antisynchronization~\cite{bib:article_Liu_PhysLettA_2006_354},
lag synchronization~\cite{bib:article_Rosenblum_PhysRevLett_1997_78},
frequency synchronization~\cite{bib:article_Anishenko_TechPhysLett_1988_6},
phase synchronization~\cite{bib:article_Pikovsky_JourBifChaos_2000_10},
and time scale synchronization~\cite{bib:article_Koronovskii_JETPL_2004_79}.
For each type, an appropriate analytic apparatus and diagnostic methods have been developed. Nevertheless, intensive investigations are being continued that are aimed, on the one hand, at the examination of different types of synchronization from unique positions and, on the other hand, at the search for new types of synchronous behavior that do not fall under the above-mentioned types. In spite of the long history of the study of synchronization of chaotic oscillations, many important problems in this field remain unsolved. These include the quantitative analysis of the time structure of synchronization of dynamical systems. By this structure we mean spikes in the synchronous behavior of the phase variables of systems between which the level of synchronism is characterized by a small parameter, i.e., intermittent behavior~\cite{bib:article_Zeldovich_UFN_1987_5}.

The concept of intermittency is very important in physics (and not only in physics) for the study of structural properties of processes and is not restricted to the synchronism of chaotic systems. This fact was noticed even by Mandelbrot~\cite{bib:article_Mandelbrot_JFluidMech_1974_2} (the turbulent flow problem
in fluid dynamics) and, somewhat later, by Zel’dovich and colleagues~\cite{bib:article_Zeldovich_UFN_1987_5} (problems of chemical kinetics and self-excitation of a magnetic field in a random flow of a conducting fluid). For instance, the phenomenon of intermittency has application in high-energy particle physics~\cite{bib:article_Dremin_UFN_1987_3}, cosmology~\cite{bib:article_Shandarin_UFN_1983_139}, in the study of rearrangements of attractors in nonlinear dynamical systems~\cite{bib:book_Brur_2003, bib:article_Gerashenko_JETP_1999_4, bib:article_Tumenev_JETP_1995_4}, and in other fields. From the physical point of view, intermittency generally implies the emergence of some structures of different scales in a
medium (for example, vortices, localized strains, temperature irregularities) that originally could be absolutely structureless on these scales. From the mathematical viewpoint, such behavior is characterized by
the presence of rare but strong peaks in the behavior of the indicator of a structure -- a certain random variable~\cite{bib:article_Zeldovich_UFN_1987_5}.

The analysis of the structure of synchronism has theoretical significance for nonlinear dynamics itself~\cite{bib:book_Brur_2003}, as well as applied value, for example, in problems
of biology and medicine~\cite{bib:article_Ghorbani_PhysRevE_2012_2, bib:article_Borisov_PhysiolHuman_2005, bib:report_Porta_ComputersinCardiology_2005}, stochastic financial mathematics~\cite{bib:article_Tino_PatternAnalysAppl_2001_4}, and so on. The paper~\cite{bib:article_Ashwin_PhysLetA_1994_2} on the analysis of the bubbling phenomenon is one of the first publications in which the authors experimentally observed intermittency between synchronous and
asynchronous behavior in a system of coupled oscillators. However, when analyzing this phenomenon, the authors did not investigate the time structure of synchronism as such. Eventually, interest of researchers in the structural phenomena arising during the synchronization of chaos grew steadily. At present, the questions of time, space, and space–time structures of synchronism are intensively studied both from theoretical
(model)~\cite{bib:article_Zueco_PhysRevE_2005_3, bib:article_Casagrande_Physica_D_2005_1, bib:article_Palaniyandi_ChaosSolitonsFractals_2008_4, bib:article_Restrepo_PhysRevE_2004_6} and applied standpoints, for example, from the viewpoint of neurophysiology (neurobiology)~\cite{bib:article_Ghorbani_PhysRevE_2012_2, bib:article_Ahn_Chaos_2013_1} and power networks~\cite{bib:article_Pecora_NatureComm_2014_6}.

For all the significance of the problems of quantitative analysis of the time structure of synchronization of nonlinear systems, progress in this field falls far behind the investigation of purely spatial synchronization patterns in distributed systems. This is clearly illustrated by the above-cited papers~\cite{bib:article_Ghorbani_PhysRevE_2012_2, bib:article_Zueco_PhysRevE_2005_3, bib:article_Casagrande_Physica_D_2005_1, bib:article_Palaniyandi_ChaosSolitonsFractals_2008_4, bib:article_Restrepo_PhysRevE_2004_6, bib:article_Ahn_Chaos_2013_1, bib:article_Pecora_NatureComm_2014_6}. Among the main reasons, in my opinion, are two objective difficulties. First, this is the lack of a unified and complete theory of nonlinear dynamical systems~\cite{bib:book_Brur_2003}. Second, during the analysis of spatial patterns, one can extensively apply both the apparatus of network theory~\cite{bib:article_Battiston_PhysRevE_2014_3} and the apparatus of the theory of pattern recognition
on images (two- and three-dimensional scalar fields of physical quantities). In turn, during the analysis of a time or space–time structure, one faces the problem of transition from time series of phase variables to the graphs or images that characterize the intermittency of the synchronization process. Not least of all, this is associated with the underdevelopment of appropriate tools, because researchers have mainly been dealing with the so-called integral coefficient of synchronism, thus studying the synchronization parameters on average. Accordingly, the measurement techniques have been developed, as a rule, within the same concepts. Note that the paradigm of studying systems and processes on average has been widely accustomed in many
experimental fields of science. The reasons for the wide use of this approach and its fundamental restrictions are presented in detail, for example, in the survey~\cite{bib:article_Zeldovich_UFN_1987_5}.

In this paper, we develop an original method for the diagnostics and quantitative measurement of the characteristics of intermittency regimes during synchronization of chaotic systems, which is aimed at the integrated study of the time structure of synchronism through the analysis of the so-called T-synchronization~\cite{bib:article_Makarenko_TechPhysLett_2012_17, bib:article_Makarenko_MatPhisModel_2013_2, bib:report_Makarenko_AFS_2013}. The method is based on the formalism of symbolic CTQ-analysis proposed by the present author~\cite{bib:article_Makarenko_TechPhysLett_2012_4, bib:article_Makarenko_CompMathMathPhys_2012_7} (the abbreviation CTQ stands for three alphabets with which the method operates: C, T, and Q). One should note that symbolic dynamics, for all its seeming external simplicity, is a very strongly substantiated tool for the analysis of nonlinear dynamical systems~\cite{bib:book_Bouen_1979} and allows one to investigate complicated phenomena in systems such as chaos, strange attractors, hyperbolicity, structural stability, controllability, etc. (see, for example, \cite{bib:book_Bouen_1979, bib:book_Hsu_1987, bib:article_Dellnitz_NumerMathem_1997_75} and references therein).

In the present study, we propose to consider the relationship between synchronous and desynchronous domains integrally. This allows us to carry out a detailed analysis of intermittency regimes during synchronization of chaotic systems. Moreover, we essentially revise the principles of the symbolic CTQ-analysis: we formulate the encoding rules for the symbols of the base alphabet in a rigorous and formal form, which allows us to form a complete set of symbols. In addition, we propose a number of measures for evaluating the time structure of synchronization. As an example, we take a system of logical mappings, which, on the one hand, is a standard object of nonlinear dynamics, and, on the other hand, with regard to the Feigenbaum universality, many results of the analysis of this system are extended to a wide class of both model and real objects.

\section{Definition of T-synchronization of chaotic systems}

Introduce a trajectory of a dynamical system defined as a discrete sequence (time series)~$\{\mathbf{s}_{k}\}^K_{k=1}$, where the phase variable~$\mathbf{s}$ of the system has dimension~$N$ and the trajectory consists of~$K$ time samples. In this
case, each $k$th sample can be assigned a time instant~$t_k$.

Define the original mapping that encodes the form of the $n$th component of the sequence in terms of a finite T-alphabet~\cite{bib:article_Makarenko_TechPhysLett_2012_4, bib:article_Makarenko_CompMathMathPhys_2012_7}:
\begin{equation}\label{eq:mapping_TSymb}
\left\lbrace\mathbf{s}^{(n)}_{k-1},\,\mathbf{s}^{(n)}_{k},\,\mathbf{s}^{(n)}_{k+1}\right\rbrace\Rightarrow
T^{\alpha\varphi}_k|_n,\quad
T^{\alpha\varphi}_k = \left[T^{\alpha\varphi}_k|_1,\,\ldots,\,T^{\alpha\varphi}_k|_N\right].
\end{equation}

Strictly speaking, mapping~(\ref{eq:mapping_TSymb}) is defined by the relations
\begin{equation}\label{eq:TSymb}
\begin{aligned}
 &\mathtt{T0}\quad   &&\Delta s_-=\Delta s_+=0,     \\
 &\mathtt{T1}\quad   &&\Delta s_-=\Delta s_+<0,     \\
 &\mathtt{T2}\quad   &&\Delta s_-=\Delta s_+>0,     \\
 &\mathtt{T3N}\quad  &&\Delta s_-<0,\quad \Delta s_+<\Delta s_-,  \\
 &\mathtt{T3P}\quad  &&\Delta s_-<0,\quad \Delta s_+<0,\quad \Delta s_+>\Delta s_-,  \\
 &\mathtt{T4N}\quad  &&\Delta s_->0,\quad \Delta s_+=0,  \\
 &\mathtt{T4P}\quad  &&\Delta s_-<0,\quad \Delta s_+=0,  \\
 &\mathtt{T5N}\quad  &&\Delta s_->0,\quad \Delta s_+>0,\quad \Delta s_+<\Delta s_-,  \\
 &\mathtt{T5P}\quad  &&\Delta s_->0,\quad \Delta s_+>\Delta s_-,  \\
 &\mathtt{T6S}\quad  &&\Delta s_->0,\quad \Delta s_+<0,\quad \Delta s_+> -\Delta s_-,  \\
 &\mathtt{T6}\quad   &&\Delta s_-=-\Delta s_+>0,  \\
 &\mathtt{T6L}\quad  &&\Delta s_->0,\quad \Delta s_+<0,\quad \Delta s_+< -\Delta s_-,  \\
 &\mathtt{T7S}\quad  &&\Delta s_-<0,\quad \Delta s_+>0,\quad \Delta s_+< -\Delta s_-,  \\
 &\mathtt{T7}\quad   &&\Delta s_-=-\Delta s_+<0,  \\
 &\mathtt{T7L}\quad  &&\Delta s_-<0,\quad \Delta s_+>0,\quad \Delta s_+> -\Delta s_-,  \\
 &\mathtt{T8N}\quad  &&\Delta s_-=0,\quad \Delta s_+<0,  \\
 &\mathtt{T8P}\quad  &&\Delta s_-=0,\quad \Delta s_+>0,
\end{aligned}
\end{equation}
where~$\Delta s_- = \mathbf{s}^{(n)}_{k}-\mathbf{s}^{(n)}_{k-1}$ и~$\Delta s_+ = \mathbf{s}^{(n)}_{k+1}-\mathbf{s}^{(n)}_{k}$.

\begin{figure}[!htb]
\begin{center}
\includegraphics[width=131mm, height=31mm]{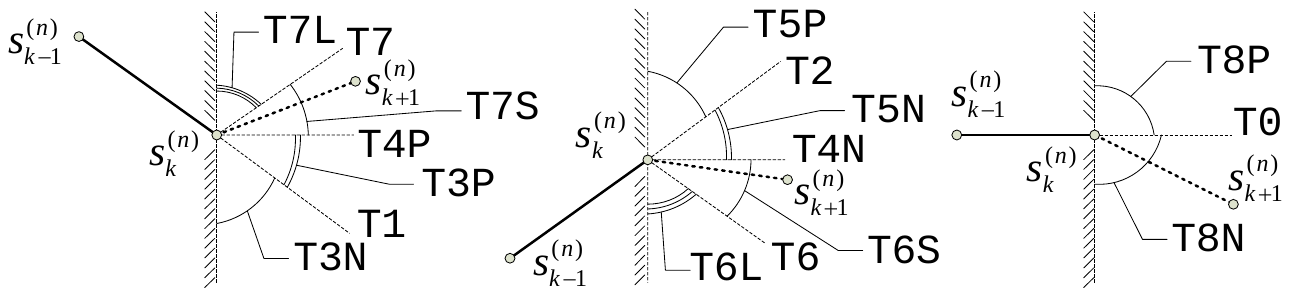}
\caption{Graphic diagrams illustrating the geometry of the symbols of~$T^{\alpha\varphi}|_n$ for the $k$th sample and the $n$th phase variable.}\label{fig:TSymb_diag}
\end{center}
\end{figure}

The graphic diagrams illustrating the geometry of the symbols~$T^{\alpha\varphi}|_n$ for the $k$th sample and $n$th phase variable are shown in Fig.~\ref{fig:TSymb_diag}.

Thus, the T-alphabet includes the set of symbols
\begin{multline} \label{eq:T_alphabet_o_n}
\mathrm{T}^{\alpha\varphi}_o=\{\mathtt{T0},\,\mathtt{T1},\,\mathtt{T2},\,\mathtt{T3N},
\,\mathtt{T3P},\,\mathtt{T4N},\,\mathtt{T4P},\,\mathtt{T5N},\,\mathtt{T5P},\,\\
\mathtt{T6S},\,\mathtt{T6},\,\mathtt{T6L},\,\mathtt{T7S},\,\mathtt{T7},\,\mathtt{T7L},
\,\mathtt{T8N},\,\mathtt{T8P}\}.
\end{multline}

This formula shows that a symbol~$T^{\alpha\varphi}_k|_n$ is encoded as~$\mathtt{T}\,i$, where~$i$ is the right part of the symbol codes of the alphabet~$\mathrm{T}^{\alpha\varphi}_o$. In turn, a symbol~$T^{\alpha\varphi}_k$ is encoded in terms of~$\mathtt{T}\,i_1\,\cdots\,i_N$, see~(\ref{eq:T_alphabet_o_n}). The complete alphabet~$\mathrm{T}^{\alpha\varphi}_o|N$ that encodes the form of the trajectory of a multidimensional sequence~$\{\mathbf{s}_k\}^K_{k=1}$ consists of~$17^N$ symbols in total.

Now, suppose, for simplicity but without loss of generality, that the time sequence~$\{\mathbf{s}_{k}\}^K_{k=1}$ of dimension~$N$ is formed by the combination of the phase variables of~$N$ one-dimensional dynamical systems; i.e., suppose that~$\mathbf{s}^{(n)}_k$ is the value of the phase variable of the~$n$th system at the~$k$th instant of time.

We will assume that the dynamical systems are synchronous at time instant~$k$ in the sense of T-synchronization~\cite{bib:article_Makarenko_TechPhysLett_2012_17} if the condition~$J_k=1$ is satisfied, where
\begin{equation} \label{eq:J_T_sync}
J_k=\begin{cases}
1 & T^{\alpha\varphi}_k|_1=\ldots=T^{\alpha\varphi}_k|_n=\ldots=T^{\alpha\varphi}_k|_N,\\
0 & \text{otherwise}.
\end{cases}\;,
\end{equation}

Taking into account possible antisynchronization~\cite{bib:article_Liu_PhysLettA_2006_354} between the systems, we should also consider all
possible variants of inversion of their phase variables: $\mathbf{s}^{(n)}_k\to -1 \cdot \mathbf{s}^{(n)}_k$. In this case, for the $n$th system, a change of symbols~$T^{\alpha\varphi}_k|_n$ of in the $k$th sample occurs by the scheme
\begin{equation} \label{eq:InvTSymb}
\begin{aligned}
&\mathtt{T0} \leftrightarrow\mathtt{T0},\\[3pt]
&\mathtt{T1} \leftrightarrow\mathtt{T2},\quad
 \mathtt{T3N}\leftrightarrow\mathtt{T5P},\quad
 \mathtt{T3P}\leftrightarrow\mathtt{T5N},\quad
 \mathtt{T4N}\leftrightarrow\mathtt{T4P},\\[3pt]
&\mathtt{T6S}\leftrightarrow\mathtt{T7S},\quad
 \mathtt{T6} \leftrightarrow\mathtt{T7},\quad
 \mathtt{T6L}\leftrightarrow\mathtt{T7L},\quad
 \mathtt{T8N}\leftrightarrow\mathtt{T8P}.
\end{aligned}
\end{equation}
Denote each variant of inversion by number~$m$. The total number of variants of inversion is~$M=2^{N-1}$.

Synchronization between systems can also be set in the lag mode~\cite{bib:article_Rosenblum_PhysRevLett_1997_78}. To detect this synchronization, one should move a little the phase trajectories of the systems with respect to each other ((shifts~$h_n\geqslant 0$):
\begin{equation} \label{eq:Shift_n}
\left\{
T^{\alpha\varphi}_k|_1 \to T^{\alpha\varphi}_{k+h_1}|_1,\, \ldots,\,
T^{\alpha\varphi}_k|_N \to T^{\alpha\varphi}_{k+h_N}|_N
\right\}.
\end{equation}

The antisynchronization and lag synchronization modes may coexist; therefore, when calculating a partial integral coefficient of synchronism, we take this fact into consideration:
\begin{equation} \label{eq:sigma_sync_h}
\delta^s_{m,\mathbf{h}} = \frac{1}{K^* + 1 - k^*}\sum\limits^{K^*}_{k = k^*} {J_k|\left\{m,\,\mathbf{h}\right\}},
\end{equation}
where $k^* = 1 + \max \left( {h_1,\, \ldots ,\,h_N} \right)$, $K^* = K + \min \left( {h_1,\, \ldots ,\,h_N} \right)$, and~$K$ is the length of the sequence~$\{T^{\alpha\varphi}_k\}^K_{k=1}$.

On the basis of the partial coefficient, we calculate the total integral coefficient of synchronism of the systems:
\begin{equation} \label{eq:sigma_sync}
\delta^s = \mathop{\max}_m \mathop{\max}_\mathbf{h} \delta^s_{m,\mathbf{h}},\quad
0\leqslant\delta^s\leqslant 1,
\end{equation}
i.e., we take a combination of shifts between the trajectories of the systems and a variant of inversion of their phase variables that, taken together, provide the maximum number of samples~$k$, satisfying the condition~$J_k=1$.

It follows from the definition of the synchronization condition~(\ref{eq:J_T_sync}) that the analyzer proposed evaluates the complete synchronization level~\cite{bib:article_Pecora_PhysRevLett_1990_64} and detects antisynchronization~\cite{bib:article_Liu_PhysLettA_2006_354} with lag synchronization~\cite{bib:article_Rosenblum_PhysRevLett_1997_78} precisely in the alphabetic representation~$\mathrm{T}^{\alpha\varphi}_o$. However, according to the definition of the geometry of the symbols of the T-alphabet~(\ref{eq:TSymb}), a complete synchronization at the level of the samples of~$T^{\alpha\varphi}_k$ is a wider phenomenon compared with the complete synchronization at the level of~$\mathbf{s}_k$ -- the samples of the sequence itself. The T-synchronism of dynamical systems (with respect to the set of phase variables~$\mathbf{s}$) is considered from the viewpoint of the shape (geometric structure) of the trajectories of the systems in the extended phase space. By the shape (geometric structure) of a trajectory of a dynamical system in the extended phase space is meant its certain invariant under uniform translations and dilations of the trajectory in the space of phase variables. Thus, in a sense,
the T-synchronization deals with the topological aspects of synchronization of dynamical systems~\cite{bib:book_Brur_2003, bib:book_Bouen_1979, bib:book_Gilmore_2002}. Hence, this opens a potential possibility for the application of the analyzer proposed to the study of generalized synchronization of chaos~\cite{bib:article_Abarbanel_PhysRevE_1996_53}.

\section{Time structure of synchronization of chaotic systems}

The quantity~$\delta^s$ introduced in~(\ref{eq:sigma_sync}) characterizes the synchronism of the systems on average over a period of~$t_K-t_1$. As mentioned in the Introduction, most investigations on the synchronization of chaos are usually restricted to this situation. However, often a researcher may be interested in the time structure of synchronization of systems. Recall that by this structure one means the spikes in the synchronous behavior of the phase variables of the systems between which the synchronism level is characterized by a small quantity, i.e., one means intermittent behavior~\cite{bib:article_Zeldovich_UFN_1987_5}.

In~\cite{bib:article_Makarenko_TechPhysLett_2012_17}, the present author introduced the concept of a synchronous domain~$\mathrm{SD}$ -- a set of samples of a time series that satisfy the condition ($\vee$ is the symbol of the logical operation OR)
\begin{equation} \label{eq:SD}
\begin{aligned}
&\mathrm{SD}_r:\,\left\lbrace
J_{k'} = 1,\; J_{k''} = 0 \vee k'' = 0,\; J_{k'''} = 0 \vee k'''= K + 1
\right\rbrace,\\[3pt]
&k'=\overline{b^{\mathrm{SD}}_r,\, b^{\mathrm{SD}}_r+L_r},\quad
 k''=b^{\mathrm{SD}}_r-1,\quad
 k'''=b^{\mathrm{SD}}_r+L^{\mathrm{SD}}_r+1,
\end{aligned}
\end{equation}
where~$b^{\mathrm{SD}}_r$, $L^{\mathrm{SD}}_r$ and~$r$ are the emergence time, the length, and the ordinal number of a synchronous domain, respectively. In this case, the following conditions are satisfied: $L^{\mathrm{SD}}_r \leqslant K$, and the total number of synchronous domains (in the original sequence) $R^{\mathrm{SD}}\leqslant (K + 1) \operatorname{div} 2$.

To quantitatively describe the structure of synchronization of systems, the author introduced in~\cite{bib:article_Makarenko_TechPhysLett_2012_17} a spectral density function of synchronous domains~$\mathrm{SD}$:
\begin{equation} \label{eq:H_SD_SS}
H^{\mathrm{SD}}\left[L\right] = \sum\limits_{r = 1}^{R^{\mathrm{SD}}} {\delta[L^{\mathrm{SD}}_r,\,{L}]},
\end{equation}
where~$\delta[\circ,\,\circ]$ is the Kronecker delta and~$L = \overline {1,\,K}$.

To analyze the degree of degeneracy of a structure of synchronous domains, we additionally define a
quantity~$E^{\mathrm{SD}}$ -- the entropy of the structure of synchronous domains (according to Shannon)~\cite{bib:article_Makarenko_MatPhisModel_2013_2}, which makes sense for~$\delta^s>0$:
\begin{equation} \label{eq:Entropy_Cond_H_SD}
E^{\mathrm{SD}} = - \sum\limits^K_{i=1}{P^{\mathrm{SD}}\left[i\right]\,\ln P^{\mathrm{SD}}\left[i\right]},\quad
P^{\mathrm{SD}}\left[L\right] =\frac{H^{\mathrm{SD}}\left[L\right]}
{\sum\limits^K_{i=1}{H^{\mathrm{SD}}\left[i\right]}}.
\end{equation}
It follows from Shannon’s entropy properties that the entropy~$E^{\mathrm{SD}}$ is minimal ($E^{\mathrm{SD}}=0$) when the spectrum~$H^{\mathrm{SD}}[L]$ is degenerate (all synchronous domains have
the same length) and maximal ($E^{\mathrm{SD}}=\hat{E}^{\mathrm{SD}}$) in the case of a uniform comb spectrum~$H^{\mathrm{SD}}[L]$ with a limit number of different lengths of synchronization domains
equal to~$\hat{W}^{\mathrm{SD}}_{cmb}$:
\begin{equation} \label{eq:Count_differ_L_SD_Max_Entropy_Cond_H_SD}
\hat{W}^{\mathrm{SD}}_{cmb} = \min\left\{
\left\lfloor\frac{\sqrt{1+8\,\delta^s\,K}-1}{2} \right\rfloor,\;
K-\delta^s\,K+1
\right\},\quad
\hat{E}^{\mathrm{SD}} = \ln \hat{W}^{\mathrm{SD}}_{cmb},
 \end{equation}
where~$\lfloor a\rfloor$ is the integer part of~$a$.

On the basis of~(\ref{eq:Entropy_Cond_H_SD}) and~(\ref{eq:Count_differ_L_SD_Max_Entropy_Cond_H_SD}), we define the relative entropy of the structure of synchronous domains:
\begin{equation} \label{eq:relative_E_Cond_H_SD}
\Delta^{\mathrm{SD}}_E=\frac{E^{\mathrm{SD}}}{\hat{E}^{\mathrm{SD}}}.
\end{equation}
It makes sense to apply the quantity~$\Delta^{\mathrm{SD}}_E$ when the researcher should compare synchronization cases that differ in the values of~$\delta^s$ and/or~$K$.

Nevertheless, for the full description of the intermittent behavior of chaotic systems during synchronization, it is obviously insufficient to study only synchronous domains~$\mathrm{SD}$. To obtain a complete and closed idea of the time structure of synchronism of dynamical systems (a complete and closed representation of the intermittency structure), in this paper we additionally introduce the concept of a {\it desynchronous domain}~$\mathrm{\overline{S}D}$ -- a set of samples of a time series satisfying the condition
\begin{equation} \label{eq:NSD}
\begin{aligned}
&\mathrm{\overline{S}D}_r:\,\left\lbrace
J_{k'} = 0,\; J_{k''} = 1 \vee k'' = 0,\; J_{k'''} = 1 \vee k'''= K + 1
\right\rbrace,\\[3pt]
&k'=\overline{b^{\mathrm{\overline{S}D}}_r,\, b^{\mathrm{\overline{S}D}}_r+L^{\mathrm{\overline{S}D}}_r},\quad
 k''=b^{\mathrm{\overline{S}D}}_r-1,\quad
 k'''=b^{\mathrm{\overline{S}D}}_r+L^{\mathrm{\overline{S}D}}_r+1,
\end{aligned}
\end{equation}
where~$b^{\mathrm{\overline{S}D}}_r$, $L^{\mathrm{\overline{S}D}}_r$ and~$r$ are the emergence time, the
length, and the ordinal number of a desynchronous domain~$\mathrm{\overline{S}D}$ , respectively.

\begin{figure}[!htb]
\begin{center}
\includegraphics[width=145mm, height=90mm]{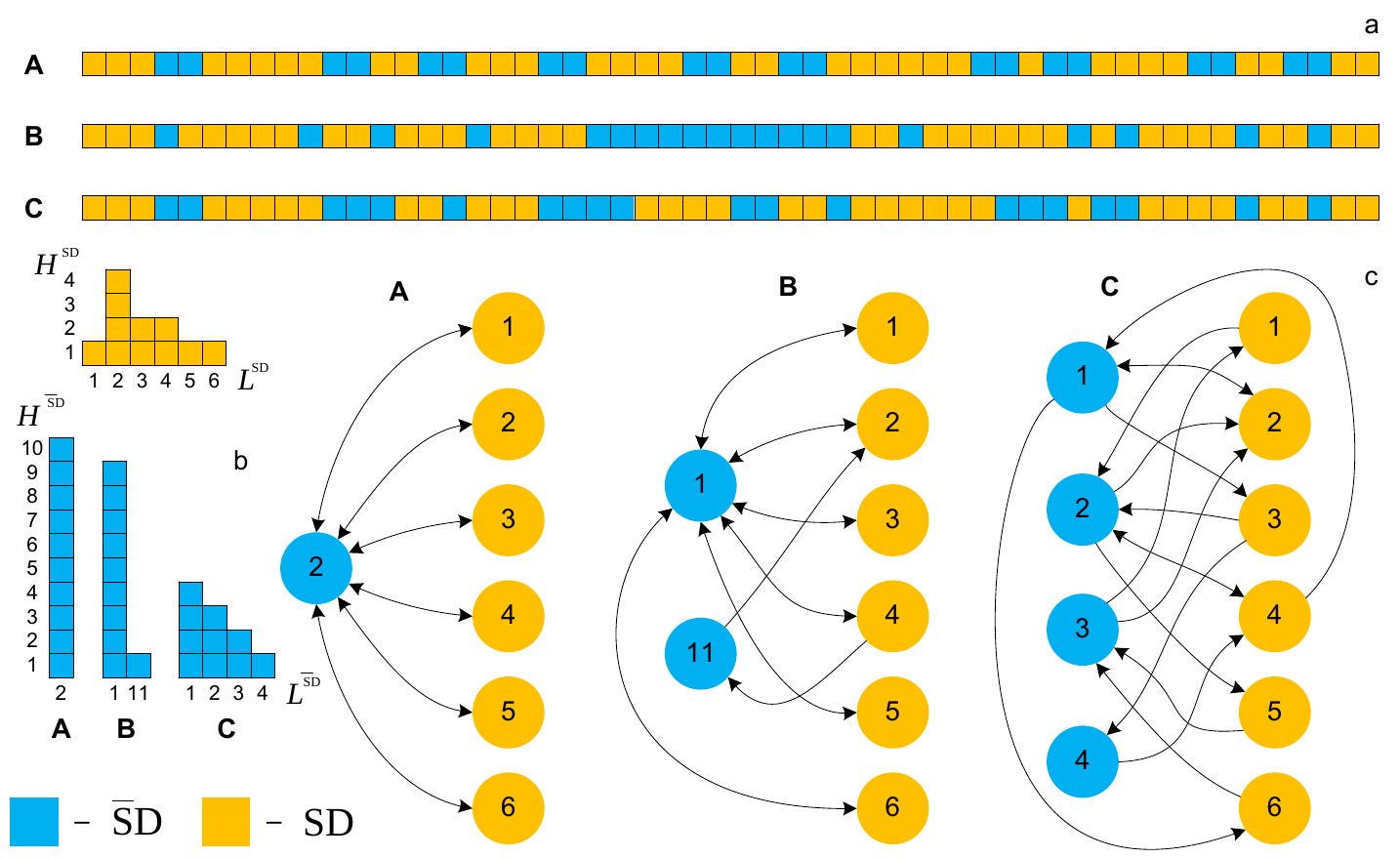}
\caption{A model example: (a) sequence order of desynchronous~$\mathrm{\overline{S}D}$ and synchronous~$\mathrm{SD}$ domains; (b) spectral densities of desynchronous and synchronous domains; and (c) scheme of transition~$\mathrm{\overline{S}D}\leftrightarrow\mathrm{SD}$ between domains of certain length. The letters A, B, and C denote three cases that have different structure of desynchronous behavior.}\label{fig:Scheme_Sync_SD}
\end{center}
\end{figure}

We will schematically illustrate the necessity of a consistent analysis of desynchronous and synchronous
domains by a model example (see Fig.~\ref{fig:Scheme_Sync_SD}a). Let us define three variants (A, B, and C) that have the same integral coefficient of synchronism, the same set and sequence of synchronous domains~$\mathrm{SD}$, but essentially different sets and sequences of desynchronous domains~$\mathrm{\overline{S}D}$.

If these model variants are compared only with respect to the parameters of~$\mathrm{SD}$ domains, then a
wrong conclusion is made on the identity of synchronization processes in these cases. However, Fig.~\ref{fig:Scheme_Sync_SD}a points to the contrary. Therefore, in addition to~$\mathrm{SD}$ domains, we consider~$\mathrm{\overline{S}D}$ domains. Figures~\ref{fig:Scheme_Sync_SD}b and~\ref{fig:Scheme_Sync_SD}c
show that, in the case of A, all desynchronous domains have a sample length of~$2$ and, in the case of B, desynchronous domains have a sample length of~$2$ and~$11$ (among which there is a single desynchronous domain of length~$11$). The case C has the most complex structure: in addition to the fact that desynchronous domains have four different lengths of~$1$, $2$, $3$, and~$4$ samples, the structure of~$\mathrm{\overline{S}D}\leftrightarrow\mathrm{SD}$ transitions is rather nontrivial. Thus, a consistent analysis of the parameters of~$\mathrm{\overline{S}D}$ and~$\mathrm{SD}$ domains, as well as of the
structure of~$\mathrm{\overline{S}D}\leftrightarrow\mathrm{SD}$ transitions, provide much more information on the character of rearrangement of the structure of attractors and on the intermittent behavior in synchronized systems. Note that, in most cases (see, for example, \cite{bib:article_Ghorbani_PhysRevE_2012_2, bib:article_Zueco_PhysRevE_2005_3, bib:article_Casagrande_Physica_D_2005_1, bib:article_Palaniyandi_ChaosSolitonsFractals_2008_4, bib:article_Restrepo_PhysRevE_2004_6, bib:article_Ahn_Chaos_2013_1, bib:article_Pecora_NatureComm_2014_6}), it is synchronization patterns that are subjected to analysis as a rule, while the desynchronization areas are considered as a background signal; i.e., they are removed from the analysis.

A quantitative analysis of the time structure of synchronization is possible through the study of a graph
(see Fig.~\ref{fig:Scheme_Sync_SD}c) whose vertices are domains~$\mathrm{\overline{S}D}$ and~$\mathrm{SD}$
with lengths~$\bar{L}$ and~$L$, respectively, and arcs (oriented edges) are~$\mathrm{\overline{S}D}\rightarrow\mathrm{SD}$ and~$\mathrm{SD}\rightarrow\mathrm{\overline{S}D}$ interdomain transitions. The labeling of the vertices of this graph is made by the probabilities~$P^{\mathrm{\overline{S}D}}$ and~$P^{\mathrm{SD}}$, and the labeling of edges, by conditional transition probabilities:
\begin{equation} \label{eq:P_SS}
P^{\mathrm{\overline{S}S}}[\bar{L},\,L] = \frac{H^{\mathrm{\overline{S}S}}[\bar{L},\,L]}
{\sum\limits^K_{\bar{l}=1}\sum\limits^K_{l=1}{H^{\mathrm{\overline{S}S}}[\bar{l},\,l]}},
\quad
P^{\mathrm{S\overline{S}}}[L,\,\bar{L}] = \frac{H^{\mathrm{S\overline{S}}}[L,\,\bar{L}]}
{\sum\limits^K_{l=1}\sum\limits^K_{\bar{l}=1}{H^{\mathrm{S\overline{S}}}[l,\,\bar{l}]}},
\end{equation}
where~$H^{\mathrm{\overline{S}S}}$ and~$H^{\mathrm{S\overline{S}}}$ -- are the spectral densities of the
interdomain transitions~$\mathrm{\overline{S}D}\rightarrow\mathrm{SD}$ and~$\mathrm{SD}\rightarrow\mathrm{\overline{S}D}$, respectively.

The analysis of the graph of the~$\mathrm{\overline{S}D}\leftrightarrow\mathrm{SD}$ transitions opens wide possibilities for the application of the currently intensively developing apparatus of network theory~\cite{bib:article_Battiston_PhysRevE_2014_3, bib:article_Davidsen_PhysRevE_2008_6, bib:article_Domenico_PhysRevX_2013_4, bib:article_Fiedor_PhysRevE_2014_5}, involving the methods of statistical physics~\cite{bib:article_Bianconi_2013_6} and symbolic dynamics~\cite{bib:book_Bouen_1979, bib:book_Hsu_1987, bib:article_Dellnitz_NumerMathem_1997_75}, to the analysis of the time structure of synchronization of chaotic oscillations. Note that these topics are the subject of our current research.

\section{Analysis of a system of logistic mappings}

A system of coupled logistic mappings is one of popular models of nonlinear dynamics for studying chaotic synchronization~\cite{bib:article_Shabunin_TechPhysLett_2001_11, bib:article_Makarenko_TechPhysLett_2012_4}. Its theoretical value is associated with the fact that, in spite of its relative simplicity, a logistic mapping gives rise to a wide spectrum of complex, including chaotic, oscillation modes~\cite{bib:article_Makarenko_CompMathMathPhys_2012_7, bib:article_Mosekilde_NonlinSci_2002_42}. A transition to such oscillation modes occurs by the classical period-doubling scenario. With regard to the Feigenbaum universality, many results are extended to a wide class of both model and real physical, biophysical, chemical, and other systems, which also arouses applied interest in the logistic mapping~\cite{bib:article_Mosekilde_NonlinSci_2002_42, bib:article_Feygenbaum_UFN_1983_2}.

Consider a system of two unidirectionally coupled logistic oscillators:
\begin{equation} \label{eq:logista_system}
\begin{aligned}
&x_{k+1} = 4\,\lambda\,x_k\,(1-x_k),\\
&y_{k+1} = 4\,\lambda\,[y_k+\gamma\,(x_k-y_k)]\,(1-[y_k+\gamma\,(x_k-y_k)]),
\end{aligned}
\end{equation}
where:~$\lambda\in(0,\,1]$ is a control parameter, $\gamma\in[0,\,1]$ is the coupling coefficient of the systems, and $x,\,y\in(0,\,1)$ are the phase variables of the master and slave systems, respectively.

The analysis of mappings~(\ref{eq:logista_system}) was carried out on the interval~ $k\in\left[1\times 10^5,\ 2\times 10^5\right]$. Such a shift from~$k=1$ is explained by the need to reduce the spurious effect of the transient process to minimum. Moreover, all the estimates of the quantities were averaged over~$1000$ realizations of the initial conditions: $x_1=\xi_1$ and~$y_1=\xi_2$, where $\xi_1,\xi_2\in\left(0,\,1\right)$ are uncorrelated uniformly distributed pseudorandom variables. This allowed us to neutralize the memory effect induced by the initial conditions on the trajectories of the processes~$x$ and~$y$. The value of the coupling parameter was varied in the interval~$\gamma\in\left[0,\ 0.5\right]$ with a step of~$1\times 10^{-4}$. The value of the control parameter was taken to be~$\lambda =0.95$, which means the existence of a developed chaos regime in~(\ref{eq:logista_system})~\cite{bib:article_Feygenbaum_UFN_1983_2}. Note that, according to the results of~\cite{bib:article_Makarenko_CompMathMathPhys_2012_7}, the final qualitative complexification of the chaotic trajectory occurs at~$\lambda=0.98805\ldots$. Nevertheless, the choice of the value of the control parameter is attributed to the necessity of mutual analysis and coordination of the results obtained in the present study with the results of~\cite{bib:article_Makarenko_TechPhysLett_2012_4, bib:article_Shabunin_TechPhysLett_2001_11}. In particular, it was established in~\cite{bib:article_Makarenko_TechPhysLett_2012_4, bib:article_Shabunin_TechPhysLett_2001_11} that, for~$\gamma\geqslant\gamma_{rb}\simeq 0.38$, a robust regime of complete synchronization is set in the system, and, at the point~$\gamma_{bb}\simeq 0.35$, a nonrobust regime of complete synchronization is set that persists on the interval~$\gamma_{bb}\leqslant\gamma<\gamma_{rb}$. There are four more points on the scale of the coupling parameter that are responsible for various rearrangements of the structure of the attractor: $\gamma_{wm}\simeq 0.0639$, $\gamma_{ws}\simeq 0.14$, $\gamma_{r1}\simeq 0.2606$ and~$\gamma_{r2}\simeq 0.2755$. For example, at~$\gamma_{ws}$, an unstable quasiperiodic motion is fixed in the attractor on which the trajectory stays for a considerable period of time~\cite{bib:article_Shabunin_TechPhysLett_2001_11}, while, at~$\gamma_{r1}$, the trajectory of the slave system~$(y)$ contains the maximum concentration of relaxation oscillations, which are missing in the master system~$(x)$~\cite{bib:article_Makarenko_TechPhysLett_2012_4}.

\begin{figure}[!htb]
\begin{center}
\includegraphics[width=160mm, height=59mm]{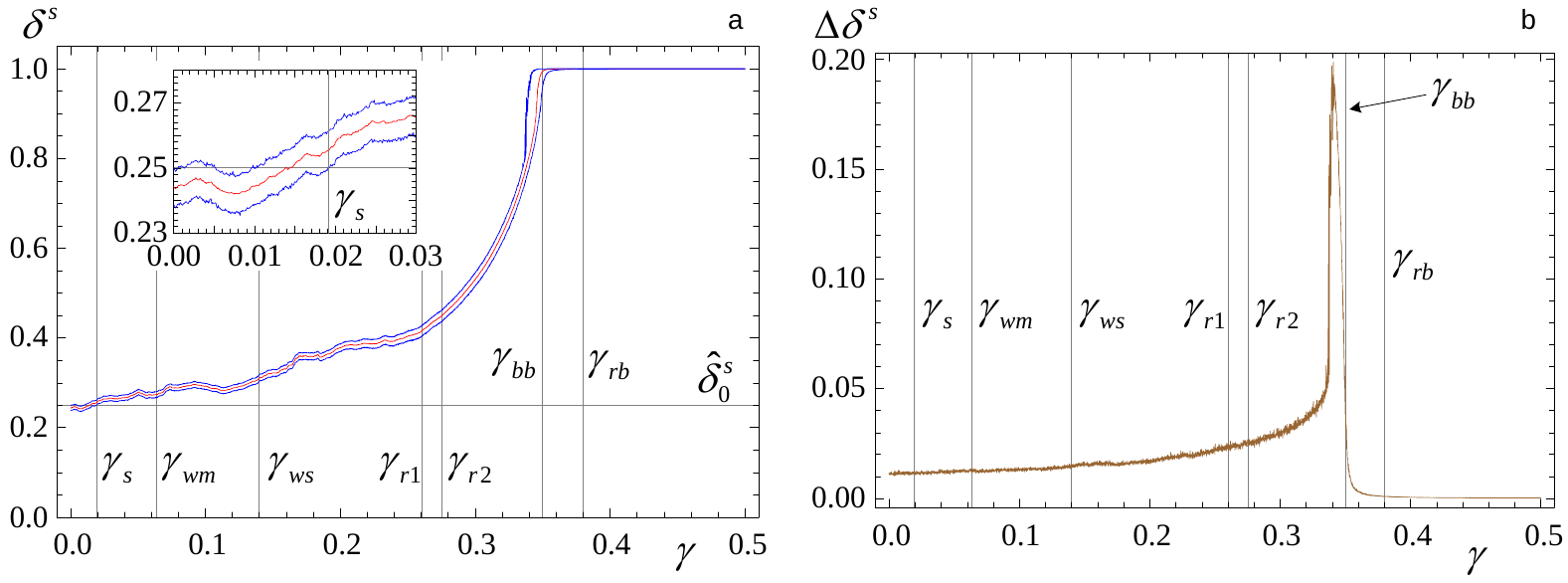}
\caption{(a) Integral level of synchronization~$\delta^s$ (the inset shows the region of small values of~$\gamma$) and (b) the width of the interval~$\Delta\delta^s = \hat{\delta}^s- \check{\delta}^s$, as a function of the coupling parameter~$\gamma$. Red curves represent an estimate for the median, and Blue curves, the interval of values with boundaries at probability~$1-\alpha$ with a significance level of~$\alpha =10^{-3}$.}\label{fig:delta_s}
\end{center}
\end{figure}

Figure~\ref{fig:delta_s}a demonstrates the graph of the integral coefficient of synchronism~$\delta^s$ as a function of the coupling parameter~$\gamma$. For~$\gamma_s\simeq 0.0191$, the lower quantile~$\check{\delta}^s$ becomes greater than the upper quantile~$\hat{\delta}^s$ for the first time. Thus, in a first approximation, the point~$\gamma_s$ can be considered as a statistically significant threshold for the emergence of T-synchronization in system~(\ref{eq:logista_system}). Note that the critical intervals were constructed by the empirical distribution functions of the calculated characteristics. The upper and lower quantiles are constructed as two-sided quantiles of orders~$1-\alpha/2$ and~$\alpha/2$, respectively. The level of statistical significance was taken to be~$\alpha =10^{-3}$. As an estimate for
the mean value of the characteristics, we used a median, as a more robust indicator compared with the
arithmetic mean~\cite{bib:book_Leman_1979}. Figure~\ref{fig:delta_s}b demonstrates the graph of~$\Delta\delta^s = \hat{\delta}^s- \check{\delta}^s$ as a function of the parameter~$\gamma$.

\begin{figure}[!htb]
\begin{center}
\includegraphics[width=159mm, height=83mm]{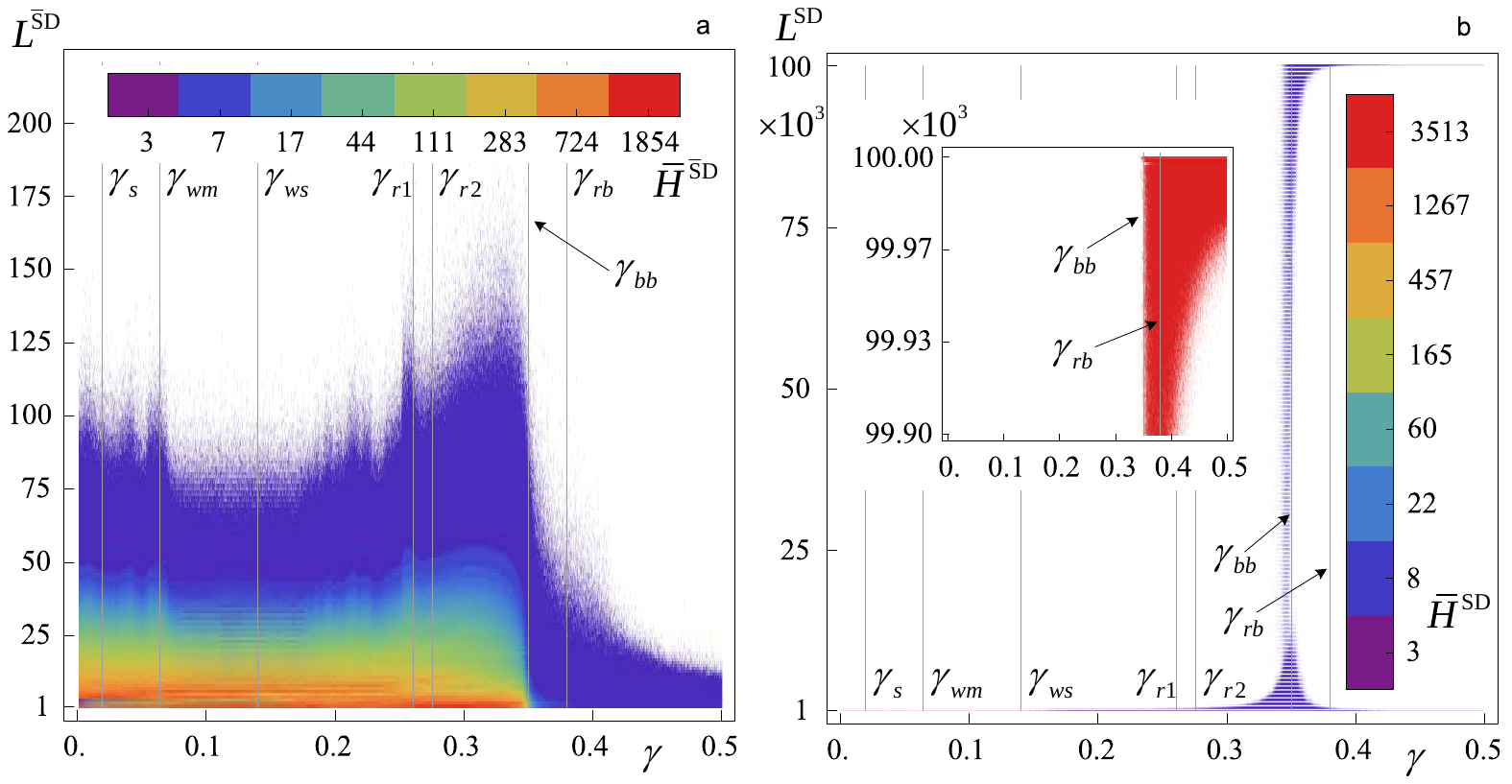}
\caption{(a) and (b) as a function of the coupling parameter~$\gamma$. The inset shows the function~$\bar{H}^\mathrm{SD}$ in the region of domains of large length~$L^\mathrm{SD}\geqslant 99900$.}\label{fig:HSS}
\end{center}
\end{figure}

Figure~\ref{fig:delta_s}a shows that the behavior of the integral coefficient of synchronism~$\delta^s$ allows one to reliably detect only a domain of bubbling behavior of the system near the point~$\gamma_{bb}$ at which a nonrobust regime of complete T-synchronization is set. The nonrobustness manifests itself in a more than fivefold stepwise broadening of the interval~$\Delta\delta^s$ (see Fig.~\ref{fig:delta_s}b). Much more complete information on the rearrangement of the structure of attractors and the time structure of synchronization is provided by the analysis of the medians of the spectral densities of the domains~$\mathrm{\overline{S}D}$ and~$\mathrm{SD}$, which are demonstrated in Fig.~\ref{fig:HSS}.

\begin{figure}[!htb]
\begin{center}
\includegraphics[width=159mm, height=83mm]{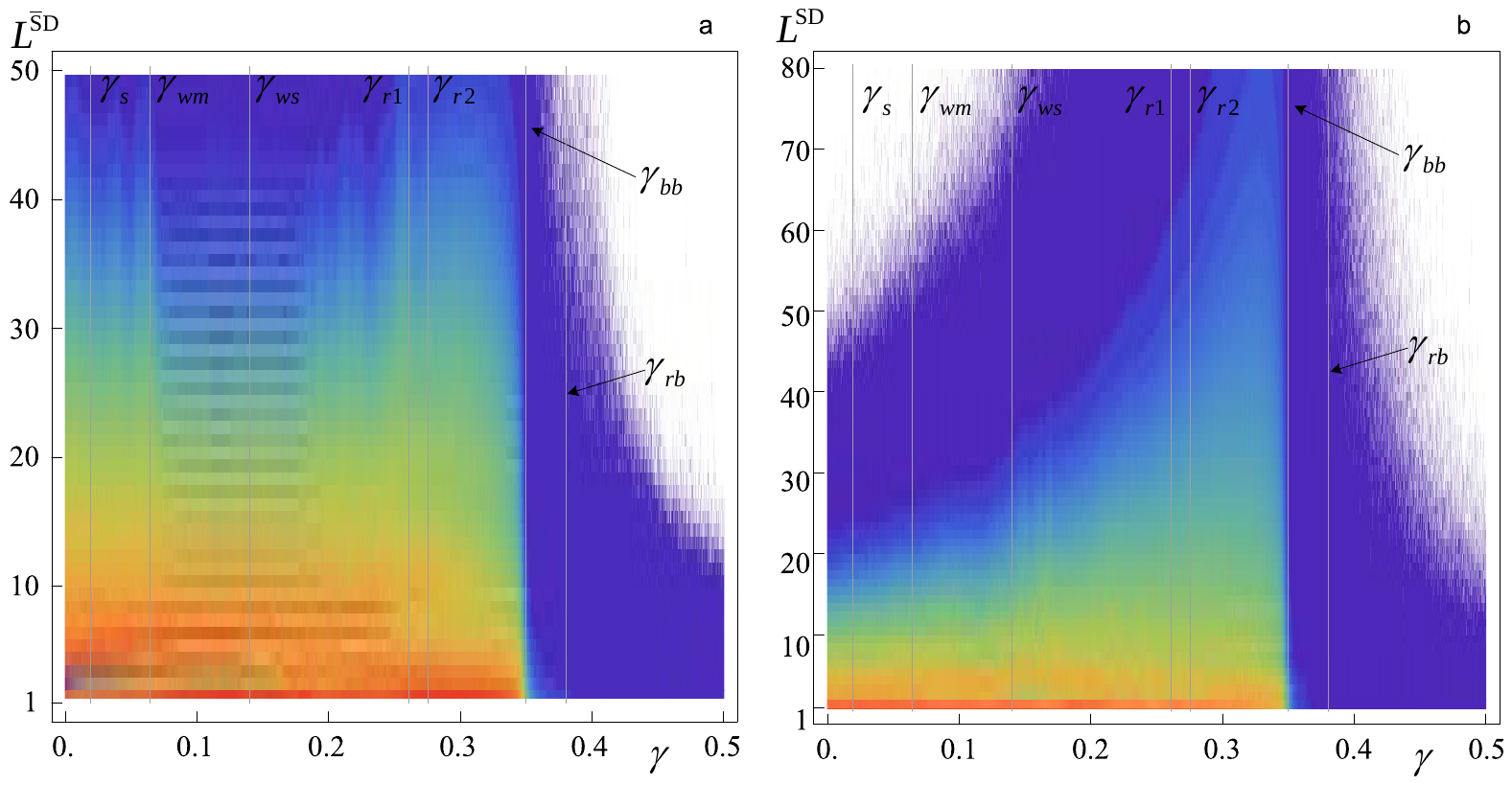}
\caption{
(a) $\bar{H}^\mathrm{\overline{S}D}$ for domains of length~$L^\mathrm{\overline{S}D}\leqslant 50$ and (b) $\bar{H}^\mathrm{SD}$ for domains of length~$L^\mathrm{SD}\leqslant 80$, as a function of the coupling parameter~$\gamma$. The color intensity scales correspond to those in Figs.~\ref{fig:HSS}a and~\ref{fig:HSS}b.}\label{fig:HSS_part}
\end{center}
\end{figure}

The time instant at which complete T-synchronization is set in system~(\ref{eq:logista_system}) at~$\gamma_{bb}$ can also be determined from the analysis of~$\bar{H}^\mathrm{SD}$ (see Fig.~\ref{fig:HSS}b). This time instant manifests itself in the burst of the lengths of synchronous domains (see the inset in Fig.~\ref{fig:HSS}b and Fig.~\ref{fig:HSS_part}b), as well as in the transition of desynchronous domains into synchronous ones.

The comparison of Figs~\ref{fig:HSS}a and~\ref{fig:HSS}b shows a significant difference between the behavior of desynchronous~$\mathrm{\overline{S}D}$ and synchronous~$\mathrm{SD}$ domains. For instance, the maximum length of synchronous domains on the entire range of variation of~$\gamma$ is not greater than~$\max L^\mathrm{\overline{S}D}=220$ samples, whereas~$\max L^\mathrm{SD}=10^5$ samples (after setting a robust regime of complete T-synchronization). As the coupling increases from~$0$ to~$\gamma_{bb}$, the lengths of synchronous domains monotonically increase, and the maximum of a spectral domain is shifted toward longer domains. This indicates that the characteristic~$\bar{H}^\mathrm{SD}$ is insensitive to the rearrangements
of the attractor structure, which occur in the slave system~$(y)$ up to the onset of the nonrobust regime of
complete T-synchronization. In turn, the analysis of~$\bar{H}^\mathrm{\overline{S}D}$ allows one to fix a number of rearrangements of the attractor structure in the system~$(y)$ that occur under the action of the master system when the coupling varies from~$0$ to~$\gamma_{bb}$. For instance, the regimes at~$\gamma_{wm}$ and~$\gamma_{r1}$ are accompanied by local spikes of the length of desynchronous domains, while the regime at~$\gamma_{ws}$ is characterized by a pronounced comblike structure of the spectrum of~$\mathrm{\overline{S}D}$ domains of small length (see Fig.~\ref{fig:HSS_part}a). This kind of spectrum indeed corresponds to an unstable quasiperiodic motion where the trajectory of system~(\ref{eq:logista_system}) stays for a significant period of time~\cite{bib:article_Shabunin_TechPhysLett_2001_11}.

\begin{figure}[!htb]
\begin{center}
\includegraphics[width=158mm, height=63mm]{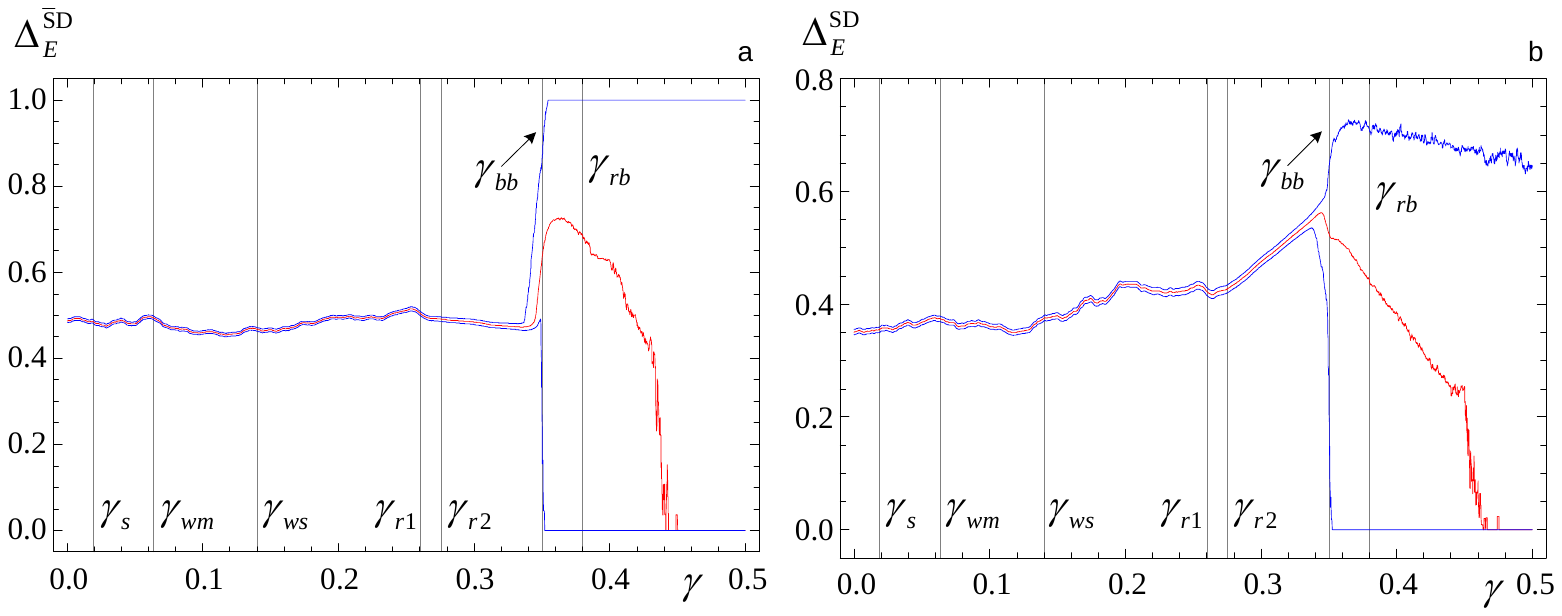}
\caption{(a)~$\Delta^{\mathrm{\overline{S}D}}_E$ and~(b)~$\Delta^{\mathrm{SD}}_E$ as a function of the coupling parameter~$\gamma$. Red curves represent an estimate for the median, and Blue curves, the interval of values with boundaries at probability~$1-\alpha$ with a significance level of~$\alpha=10^{-3}$.}\label{fig:Delta_E_SS}
\end{center}
\end{figure}

The comparison of Figs.~\ref{fig:Delta_E_SS}a and~\ref{fig:Delta_E_SS}b also demonstrates a difference in the behavior of the degree of degeneracy of the structure of synchronous and desynchronous domains, depending on~$\gamma$ -- the coupling between the master and slave systems. It follows from the analysis of the figures that, almost up to the setting a nonrobust regime of complete T-synchronization, the relative entropy of the structure of domains~$\mathrm{\overline{S}D}$ varies very little, remaining near the level of~$0.5$. There are local statistically significant spikes near the rearrangement areas of the attractor structure~$\gamma_{wm}$ and~$\gamma_{r1}$. The width of the interval of values with boundaries at probability~$1-\alpha$ with the significance level~$\alpha=10^{-3}$ also remains virtually constant. As the parameter~$\gamma$ approaches the critical value~$\gamma_{bb}$ and the system passes to a nonrobust synchronization regime, the interval of values of~$\Delta^{\mathrm{\overline{S}D}}_E$ with boundaries at probability of~$1-\alpha$ opens up, and the median~$\bar{\Delta}^{\mathrm{\overline{S}D}}_E$ experiences a jump from~$0.5$ to~$0.73$. A further increase in the coupling force to~$\gamma=0.5$, the width of the interval is not changed, while the median~$\bar{\Delta}^{\mathrm{\overline{S}D}}_E$ decreases to values close to zero.
In turn, as the coupling parameter increases from~$\gamma=0$, the relative entropy of the structure of synchronous domains~$\mathrm{SD}$ first locally (statistically significantly) increases from~$0.35$ to~$0.375$ near the value of~$\gamma_{wm}$ and then returns to the original level, and, starting from about~$\gamma_{ws}$, again starts to increase. Before the~$\gamma_{r1}$region, there is a local maximum, and, in the region~$[\gamma_{r1},\,\gamma_{r2}]$, there is a local minimum. As the coupling force increases further, the interval of values of~$\Delta^{\mathrm{SD}}_E$ with boundaries at probability of~$1-\alpha$ opens up near~$\gamma_{bb}$, and the median~$\bar{\Delta}^{\mathrm{SD}}_E$, upon reaching a maximum of~$0.57$, decreases to values close to zero.

The analysis carried out implies that the relative entropies~$\Delta^{\mathrm{\overline{S}D}}_E$ and~$\Delta^{\mathrm{SD}}_E$ are a very sensitive indicator of rearrangements of the attractor structure that occur in the regime of chaos synchronization.

\section{Conclusions}

We have proposed a new method for integrated analysis of the time structure of synchronization in nonlinear chaotic systems. The method allows one to diagnose and quantitatively evaluate the intermittency characteristics during synchronization of chaotic oscillations in the so-called T-synchronization mode~\cite{bib:article_Makarenko_TechPhysLett_2012_17, bib:article_Makarenko_MatPhisModel_2013_2}. The approach is based on the formalism of symbolic CTQ-analysis proposed by the author in~\cite{bib:article_Makarenko_TechPhysLett_2012_4, bib:article_Makarenko_CompMathMathPhys_2012_7}.

To demonstrate the main capabilities of the method, we have carried out an analysis of a system of two identical logistic oscillators with unidirectional coupling that are in the developed chaos regime. This system is a standard model of nonlinear dynamics. In addition, in view of the Feigenbaum universality, many results of the analysis can be extended to a wide class of both theoretical and real objects~\cite{bib:article_Mosekilde_NonlinSci_2002_42, bib:article_Feygenbaum_UFN_1983_2}. The analysis carried out has revealed the setting of nonrobust and robust regimes of complete T-synchronization in the system. The nonrobust regime is accompanied by a bubbling attractor that exhibits intermittency between synchronous and asynchronous behavior of the oscillators. In addition, we have confirmed the results of~\cite{bib:article_Shabunin_TechPhysLett_2001_11} concerning the fact that, for certain values of the coupling force between subsystems, an unstable quasiperiodic motion is detected in the attractor on which the trajectory stays for a considerable period of time. We have also established that the regime, revealed in~\cite{bib:article_Makarenko_TechPhysLett_2012_4}, in which the trajectory of the slave subsystem contains the maximum concentration of relaxation oscillations, which are missing in the master subsystem, leads to a local increase in the deviation of the length of desynchronous domains. This fact increases the intermittency between the synchronous and asynchronous behavior of the oscillators.

The analysis has confirmed that the widely used method in which only synchronization patterns are analyzed, while the desynchronization areas are considered as a background signal and are actually not analyzed, should be regarded as methodologically incomplete due to the loss of important information on the character of rearrangement of the structure of attractors and the intermittent behavior in synchronizing systems.

In conclusion, note that the method considered, which is based on the analysis of T-synchronization, can be successfully applied to the study of multidimensional systems consisting of two or a greater number of coupled nonidentical oscillators, including multidimensional lattices of oscillators with arbitrary topology. The approach described can be applied to the analysis of experimental data, because it does not require any a priori knowledge of a system under study. Moreover, the invariance of the analyzer with respect to the shifts and dilations of phase trajectories~\cite{bib:article_Makarenko_CompMathMathPhys_2012_7} allows one to investigate the synchronization of strongly nonstationary systems. It is quite possible that this fact will allow one to effectively apply the method proposed to the analysis of multidimensional time series generated by physical and engineering~\cite{bib:book_Pikovsky_2001, bib:article_Kuznetsov_UFN_2011_2, bib:article_Napartovich_JETP_1999_5}, biophysical~\cite{bib:article_Ghorbani_PhysRevE_2012_2, bib:article_Borisov_PhysiolHuman_2005, bib:report_Porta_ComputersinCardiology_2005}, financial~\cite{bib:article_Tino_PatternAnalysAppl_2001_4, bib:report_Makarenko_AFS_2013, bib:article_Fiedor_PhysRevE_2014_5}, power~\cite{bib:article_Pecora_NatureComm_2014_6}, and other systems.

\begin{Biblioen}

\bibitem{bib:book_Pikovsky_2001}
{\it A.S. Pikovsky, M.G. Rosenblum, J. Kurths}, Synchronization: a universal concept in nonlinear sciences. Cambridge University Press, Cambridge (2001).

\bibitem{bib:article_Boccaletti_PhysRep_2002_366}
{\it J.~Kurths, G. V.~Osipov, D. L.~Valladares, and C.S.~Zhou}, Physics Reports {\bf 366}, 1 (2002).

\bibitem{bib:article_Argonov_JETPL_2004_80}
{\it V.Yu. Argonov and S.V. Prants}, JETP Lett. {\bf 80}~(4), 231 (2004).

\bibitem{bib:article_Kuznetsov_UFN_2011_2}
{\it S.P. Kuznetsov}, Phys.--Usp. {\bf 54}~(2), 119 (2011).

\bibitem{bib:article_Napartovich_JETP_1999_5}
{\it A.P. Napartovich and A.G. Sukharev}, J.~Exp.~Theor.~Phys. {\bf 88}~(5), 875 (1999).

\bibitem{bib:article_Cuomo_PhysRevLett_1993_71}
{\it K.M. Cuomo and A.V. Oppenheim}, Phys. Rev. Lett. {\bf 71}, 65 (1993).

\bibitem{bib:article_Larger_Physique_2004_5}
{\it L. Larger and J.-P. Goedgebuer}, C.R. Physique {\bf 5}, 609 (2004).

\bibitem{bib:article_Planat_Neuroquantology_2004_2}
{\it M. Planat}, Neuroquantology {\bf 2}, 292 (2004); arXiv~quant-ph/0403020.

\bibitem{bib:article_Abarbanel_PhysRevE_1996_53}
{\it H.D.I. Abarbanel, N.F. Rulkov, and M.M. Sushchik}, Phys. Rev. E {\bf 53}, 4528 (1996).

\bibitem{bib:article_Pecora_PhysRevLett_1990_64}
{\it L.M. Pecora and T.L. Caroll}, Phys. Rev. Lett. {\bf 64}, 821 (1990).

\bibitem{bib:article_Liu_PhysLettA_2006_354}
{\it W.~Liu, X.~Qian, J.~Yang, and J.~Xiao}, Phys. Lett. A {\bf 354}, 119 (2006).

\bibitem{bib:article_Rosenblum_PhysRevLett_1997_78}
{\it M.G.~Rosenblum, A.S.~Pikovsky, and J.~Kurths}, Phys. Rev.Lett. {\bf 78}, 4193 (1997).

\bibitem{bib:article_Anishenko_TechPhysLett_1988_6}
{\it V.S.~Anishchenko and D.E.~Postnov}, Sov. Tech. Phys. Lett. {\bf 14}~(3), 254 (1988).

\bibitem{bib:article_Pikovsky_JourBifChaos_2000_10}
{\it A.S. Pikovsky, M.G. Rosenblum, and J. Kurths}, Int. J. of Bifurcation and Chaos {\bf 10}, 2291 (2000).

\bibitem{bib:article_Koronovskii_JETPL_2004_79}
{\it A.A. Koronovskii and A.E. Khramov}, JETP Lett. {\bf 79}~(7), 316 (2004).

\bibitem{bib:article_Zeldovich_UFN_1987_5}
{\it Ya.B.~Zel’dovich, S.A.~Molchanov, A.A.~Ruzmaikin, and D.D.~Sokolov}, Sov. Phys.--Usp. {\bf 30}~(5), 353
(1987).

\bibitem{bib:article_Mandelbrot_JFluidMech_1974_2}
{\it B.B. Mandelbrot}, J. Fluid Mech. {\bf 62}, 331 (1974).

\bibitem{bib:article_Dremin_UFN_1987_3}
{\it I.M. Dremin}, Sov. Phys.--Usp. {\bf 30}~(7), 649 (1987).

\bibitem{bib:article_Shandarin_UFN_1983_139}
{\it S.F.~Shandarin, A.G.~Doroshkevich, and Ya.B.~Zel’dovich}, Sov. Phys.--Usp. {\bf 26}~(1), 46 (1983).

\bibitem{bib:book_Brur_2003}
{\it H.W.~Broer, F.~Dumortier, S.J.~van~Strien, and F.~Takens}, Structure in Dynamics: Finite Dimensional Deterministic Studies (Studies in Mathematical Physics). North-Holland, Amsterdam, The Netherlands (1991).

\bibitem{bib:article_Gerashenko_JETP_1999_4}
{\it O.V. Gerashchenko}, J. Exp. Theor. Phys. {\bf 89}~(4), 797 (1999).

\bibitem{bib:article_Tumenev_JETP_1995_4}
{\it V.K. Tyumenev}, J. Exp. Theor. Phys. {\bf 80}~(4), 754 (1995).

\bibitem{bib:article_Ghorbani_PhysRevE_2012_2}
{\it M.~Ghorbani, M.~Mehta, R.~Bruinsma, and A.J.~Levine}, Phys. Rev. E {\bf 85}, 021908 (2012).

\bibitem{bib:article_Borisov_PhysiolHuman_2005}
{\it S.V.~Borisov, A.Ya.~Kaplan, N.L.~Gorbachevskaya, and I.A.~Kozlova}, Hum. Physiol. {\bf 31}~(3), 255 (2005).

\bibitem{bib:report_Porta_ComputersinCardiology_2005}
{\it A.~Porta, G.~D’Addio, G.D.~Pinna, R.~Maestri, T.~Gnecchi-Ruscone, R.~Furlan, N.~Montano, S.~Guzzetti, and A.~Malliani}, Symbolic analysis of~24h holter heart period variability series: comparison between normal and heart failure patients, Proceedings Computers in Cardiology 2005, 575 (2005).

\bibitem{bib:article_Tino_PatternAnalysAppl_2001_4}
{\it P.~Tino, C.~Schittenkopf, and G.~Dorffner}, Pattern Analysis \& Appl. {\bf 4} 283 (2001).

\bibitem{bib:article_Ashwin_PhysLetA_1994_2}
{\it P.~Ashwin, J.~Buescu, and I.~Stewart}, Physics Letters A {\bf 193}, 126 (1994).

\bibitem{bib:article_Zueco_PhysRevE_2005_3}
{\it D.~Zueco, P.J.~Martinez, L.M.~Floria, and F.~Falo}, Phys. Rev. E {\bf 71}, 036613 (2005).

\bibitem{bib:article_Casagrande_Physica_D_2005_1}
{\it V.~Casagrande and A.S.~Mikhailov}, Physica D {\bf 205}, 154 (2005).

\bibitem{bib:article_Palaniyandi_ChaosSolitonsFractals_2008_4}
{\it P.~Palaniyandi, P.~Muruganandam, and M.~Lakshmanan}, Chaos, Solitons and Fractals {\bf 36}, 991 (2008).

\bibitem{bib:article_Restrepo_PhysRevE_2004_6}
{\it J.G.~Restrepo, E.~Ott, and B.R.~Hunt}, Phys. Rev. E 69, 066215 (2004)

\bibitem{bib:article_Ahn_Chaos_2013_1}
{\it S.~Ahn and L.L.~Rubchinsky}, Chaos {\bf 23}, 013138 (2013).

\bibitem{bib:article_Pecora_NatureComm_2014_6}
{\it L.M.~Pecora, F.~Sorrentino, A.M.~Hagerstrom, T.E.~Murphy, and R.~Roy}, Nature Comm. {\bf 5}, 4079 (2014).

\bibitem{bib:article_Battiston_PhysRevE_2014_3}
{\it F.~Battiston, V.~Nicosia, and V.~Latora}, Phys. Rev. E {\bf 89}, 032804 (2014).

\bibitem{bib:article_Makarenko_TechPhysLett_2012_17}
{\it A.V. Makarenko}, Tech. Phys. Lett. {\bf 38}~(17), 804 (2012); arXiv:1212.2724.

\bibitem{bib:article_Makarenko_MatPhisModel_2013_2}
{\it A.V. Makarenko}, Nanostrukt.: Mat. Fiz. Model. {\bf 8}, 21 (2013).

\bibitem{bib:report_Makarenko_AFS_2013}
{\it A.V. Makarenko}, Symbolic CTQ-analysis –- a new method for studying of financial indicators, in Abstracts of Papers of the International Conference "Advanced Finance and Stochastics", Steklov Mathematical Institute, Moscow, Russia, June~24-28, 2013, 63 (2013).

\bibitem{bib:article_Makarenko_TechPhysLett_2012_4}
{\it A.V. Makarenko}, Tech. Phys. Lett. {\bf 38}~(2), 155 (2012); arXiv:1203.4214.

\bibitem{bib:article_Makarenko_CompMathMathPhys_2012_7}
{\it A.V. Makarenko}, Comp. Math. Math. Phys. {\bf 52}~(7), 1017 (2012).

\bibitem{bib:book_Bouen_1979}
{\it R.~Bowen}, Amer. J. Math. {\bf 95}~(429), 459 (1973).

\bibitem{bib:book_Hsu_1987}
{\it C.S.~Hsu}, Cell-to-Cell Mapping: A method of Global Analysis for Nonlinear Systems, Springer-Verlag, N.Y. (1987).

\bibitem{bib:article_Dellnitz_NumerMathem_1997_75}
{\it M.~Dellnitz and A.~Hohmann}, Numerische Mathematik {\bf 75}, 293 (1997).

\bibitem{bib:book_Gilmore_2002}
{\it R. Gilmore and M. Lefranc}, The topology of chaos. Wiley-Interscience, New York (2002).

\bibitem{bib:article_Davidsen_PhysRevE_2008_6}
{\it J.~Davidsen, P.~Grassberger, and M.~Paczuski}, Phys. Rev. E {\bf 77}, 066104 (2008).

\bibitem{bib:article_Domenico_PhysRevX_2013_4}
{\it M.~Domenico, A.~Sole-Ribalta, E.~Cozzo, M.~Kivela, Y.~Moreno, M.A.~Porter, S.~Gomez, and A.~Arenas}, Phys. Rev. X {\bf 3}, 041022 (2013).

\bibitem{bib:article_Fiedor_PhysRevE_2014_5}
{\it P. Fiedor}, Phys. Rev. E {\bf 89}, 052801 (2014).

\bibitem{bib:article_Bianconi_2013_6}
{\it G.~Bianconi}, Phys. Rev. E {\bf 87}, 062806 (2013).

\bibitem{bib:article_Shabunin_TechPhysLett_2001_11}
{\it A.V.~Shabunin, V.V.~Demidov, V.V.~Astakhov, and V.S.~Anishchenko}, Tech. Phys. Lett. {\bf 27}~(6), 476
(2001).

\bibitem{bib:article_Mosekilde_NonlinSci_2002_42}
{\it E.~Mosekilde, Yu.~Maistrenko, and D.~Postnov}, Chaotic synchronization: Applications to living systems.
(World Sci. Ser. Nonlinear Sci. Ser. A Monogr. Treatises, vol. 42.) River Edge, N.J.: World Sci. Publ. (2002).

\bibitem{bib:article_Feygenbaum_UFN_1983_2}
{\it M.J.~Feigenbaum}, Los Alamos Sci. {\bf 1}, 4 (1980).

\bibitem{bib:book_Leman_1979}
{\it E.L. Lehmann}, Testing Statistical Hypotheses. John Wiley and Sons, New York (1959).

\end{Biblioen}

\hfill{\it Translated by I.~Nikitin}


\noindent
\\\textsf{\textbf{Andrey V. Makarenko} -- was born in~1977, since~2002 -- Ph.~D. of Cybernetics. Founder and leader of the Research \& Development group "Constructive Cybernetics". Author and coauthor of more than 60~scientific articles and reports. Member~IEEE (IEEE Signal Processing Society Membership; IEEE Computational Intelligence Society Membership). Research interests: Analysis of the structure dynamic processes, predictability; Detection, classification and diagnosis is not fully observed objects (patterns); Synchronization and self-organization in nonlinear and chaotic systems; System analysis and math.~modeling of economic, financial, social and bio-physical systems and processes; Convergence of Data~Science, Nonlinear~Dynamics and~Network-Centric.}

\end{document}